\documentclass[apj]{emulateapj}
\usepackage{amssymb}
\usepackage{graphicx}

\def\lea{\mathrel{<\kern-1.0em\lower0.9ex\hbox{$\sim$}}}
\def\gea{\mathrel{>\kern-1.0em\lower0.9ex\hbox{$\sim$}}}

\slugcomment{To appear in ApJ, 2011} 
\shorttitle{Extended Star Clusters in NGC 6822} \shortauthors{Hwang et al.}

\begin{document}

\title{Extended Star Clusters in the Remote Halo of the Intriguing Dwarf Galaxy NGC
6822\altaffilmark{*}}

\author{Narae Hwang\altaffilmark{1}, Myung Gyoon Lee\altaffilmark{2},
Jong Chul Lee\altaffilmark{2}, Won-Kee Park\altaffilmark{3}, \\
Hong Soo Park\altaffilmark{2,4}, Sang Chul Kim\altaffilmark{4}, and Jang-Hyun
Park\altaffilmark{4}}

\altaffiltext{*}{Based on observations obtained with MegaPrime/MegaCam, a joint
project of CFHT and CEA/DAPNIA, at the Canada-France-Hawaii Telescope (CFHT) which is
operated by the National Research Council (NRC) of Canada, the Institut National des
Science de l'Univers of the Centre National de la Recherche Scientifique (CNRS) of
France, and the University of Hawaii.}
\altaffiltext{1}{National Astronomical Observatory of Japan \\2-21-1 Osawa
Mitaka, Tokyo 181-8588, Japan [e-mail: narae.hwang@nao.ac.jp]}
\altaffiltext{2}{Astronomy Program, Department of Physics and Astronomy,\\
Seoul National University, Seoul 151-747, Korea}
\altaffiltext{3}{CEOU, Department of Physics and Astronomy,\\
Seoul National University, Seoul 151-747, Korea}
\altaffiltext{4}{Korea Astronomy and Space Science Institute, 61-1, Hwaam-Dong,
Yuseong-Gu, Daejeon 305-348, Korea}

\begin{abstract}
We present a study on four new star clusters discovered in the halo of
the intriguing dwarf
irregular galaxy NGC 6822 from a wide field survey covering 3$\degr \times$
3$\degr$ area carried out with MegaCam at Canada-France-Hawaii Telescope
(CFHT). The star clusters have extended structures with half-light radii
${R}_{\rm h} \approx 7.5$--$14.0$ pc, larger than typical Galactic globular clusters
and other known globular clusters in NGC 6822. The integrated colors
and color magnitude diagrams (CMD) of resolved stars suggest that the new star
clusters
are 2 -- 10 Gyr old and relatively metal poor with Z=0.0001--0.004
based on the comparison with theoretical models. The projected distance of
each star cluster from the galaxy center ranges from $10.7\arcmin$
($\approx 1.5$ kpc) to $77\arcmin$ ($\approx 11$ kpc), far beyond the
optical body of the galaxy. Interestingly, the new star clusters are
aligned along the elongated old stellar halo of NGC 6822, which
is almost perpendicular to the HI gas distribution where young stellar populations
exist.
We also find that the colors and half-light radii of the new clusters are correlated
with the galactocentric distance: clusters farther from the galaxy
center are larger and bluer than those closer to the galaxy center. We discuss the
stellar structure and evolution of NGC 6822 implied by these new extended
star clusters in the halo. We also discuss the current status of observational
and theoretical understandings regarding the origin of extended star clusters
in NGC 6822 and other galaxies.

\end{abstract}



\section{Introduction}

Globular clusters (GCs) are a major tracer and a component of stellar
populations in galaxies due to their well defined photometric, chemical,
and physical properties. Their color is mostly $0.5 \la (V-I) \la 1.0$,
their metallicity is [Fe/H]$\approx -2.0 \sim -1.0$, and their half-light
radius is usually $R_{\rm h} = 2 \sim 10$ pc \citep{har96}. These
properties of GCs are known to be universal regardless of the morphological
type of galaxies.

Advancement of observational studies, however, have revealed the existence
of `unusual' GC populations. Some examples include extremely luminous and
large GCs such as $\omega$Cen \citep{lee99} and faint fuzzy clusters
\citep{lar00} that are relatively faint and red but systematically larger
than typical GCs. The existence of these `peculiar' GCs has been suspected
to be related with dynamical evolutions of their host galaxies.

Recently, another new population of star clusters was discovered in nearby
galaxies. \citet{hux05} reported the discovery of three new star clusters in
M31 and showed that these star clusters are globular but systematically larger
than normal GCs with their half light radii $R_{\rm h} $ of $26 \sim 34$ pc.
Similar star clusters were also discovered in M33 \citep{sto08,coc11}.
The number of these `extended star clusters (ESC)'
in M31 has increased to 13 through the wide field survey in the halo of M31
\citep{hux08, hux11}.
A recent report of one ESC in the Sculptor Group dwarf spheroidal galaxy
by \citet{dac09} shows that ESCs are not limited to spiral galaxies but
they exist in dwarf galaxies.

NGC 6822 is a small barred dwarf irregular galaxy in the Local Group without any
associated neighbors. Stellar populations in NGC 6822 are mostly
blue stars younger than $\sim 200$ Myr \citep{gal96c}.
Many previous studies (e.g., \citealt{hod77, KH04}) reported that star clusters
are concentrated within $6\arcmin$ from the center of NGC 6822.
These include a metal poor and old GC Hubble VII (H VII) with
age $>10$ Gyr and [Fe/H]=$-1.95 \pm 0.15$
\citep{coh98}. Another GC in NGC
6822 that is of our interest is an intermediate age cluster Hubble VIII (H VIII)
with diffuse morphology \citep{wyd00}.
H VIII is known to be about $1.5 \sim 4$ Gyr old and metal poor with [Fe/H]=$-1.58
\pm 0.28$ \citep{cha00, str03}.

One more interesting point with NGC 6822 is that the main galaxy body is
enveloped with a huge HI gas cloud \citep{rob72} and the HI gas cloud
actually constitutes a disk-like rotating structure \citep{deb00,wel03}. It is
also found that young blue stars are distributed along the HI gas
structure \citep{komi03,deb03}, suggesting that the HI gas has been used for
the recent star formation. However, \citet{lee05} show that there is a
giant stellar halo composed of red giant branch (RGB) stars older than 1 Gyr
around the main body of NGC 6822. The spatial distribution of these RGB stars
indicates that the old halo is elongated with its major axis nearly
perpendicular to the HI gas structure. This is consistent with NGC 6822 halo or
spheroid composed of intermediate age C stars reported by \citet{dem06}.
These recent developments reveal that NGC 6822 is a rather complex
system even though it is a small and isolated dwarf galaxy.

In this paper, we present the result of star cluster survey made over
the halo of NGC 6822 and the photometric properties of those new star clusters.
Preliminary results of this study were given in \citet{hwa05}, introducing
three ESCs discovered in NGC 6822. We compare various
properties of these new clusters with those of two GCs in NGC 6822,
H VII and H VIII. We also discuss any correlation between these intriguing
ESCs and the large stellar halo of NGC 6822 reported by
\citet{lee05} and \citet{bat06}.
In this study, we adopt $(m-M)_0 = 23.35 \pm 0.02$ for NGC 6822,
an average of distance moduli measured by using RR Ryrae \citep{cle03},
Cepheids \citep{pie04}, and NIR TRGB \citep{cio05}.
This translates to 470 kpc in physical distance to NGC 6822 where $1\arcsec$
corresponds to about 2.2 pc in projected scale.


\begin{deluxetable*}{lcccccccccc}
\tabletypesize{\scriptsize}
\tablecaption{Coordinates and photometric information on the new star clusters,
   Hubble VII, and Hubble VIII.\label{tab1}}
\tablewidth{0pt}
\tablehead{ \colhead{ID} & \colhead{RA (J2000)} & \colhead{Dec (J2000)} &
\colhead{$g'$} & \colhead{$(g'-i')$} &
 \colhead{$E(g'-i')$} & \colhead{$A_{g'}$} & \colhead{$A_{i'}$} & \colhead{$R_{\rm
h}$}
& \colhead{$M_{\rm V,0}$} & \colhead{$(V-I)_{0}$} \\
 & \colhead{[h  m  s]} & \colhead{[d  m  s]} & \colhead{[mag]} & \colhead{[mag]} &
\colhead{[mag]}
& \colhead{[mag]} & \colhead{[mag]} & \colhead{[pc]} & \colhead{[mag]} &
\colhead{[mag]} }
 \startdata
 NGC6822C1 & 19 40 11.77  & -15 21 47.3  & 16.35 $\pm$ 0.01 & 0.93 & 0.16 & 0.36 &
0.20 & 14.0 $\pm$ 0.2 & -7.70 & 0.85
\\
 NGC6822C2 & 19 43 04.39  & -14 58 21.5  & 18.25 $\pm$ 0.01 & 1.17 & 0.27 & 0.60 &
0.33 & 11.5 $\pm$ 0.2 & -6.10 & 0.94
\\
 NGC6822C3 & 19 45 40.15  & -14 49 25.0  & 19.23 $\pm$ 0.01 & 1.67 & 0.23 & 0.51 &
0.28 & ~7.5 $\pm$ 0.5 & -5.22 & 1.31
\\
 NGC6822C4 & 19 47 30.54  & -14 26 49.3  & 18.16 $\pm$ 0.02 & 1.33 & 0.17 & 0.38 &
0.21 & 13.8 $\pm$ 0.3 & -6.06 & 1.12
\\
 Hubble VII      & 19 44 55.77  & -14 48 56.2  &  -- \tablenotemark{1}   &  --
\tablenotemark{1}  & 0.32 & 0.71 & 0.39 & ~2.5 $\pm$ 0.1 & -8.16\tablenotemark{2} &
1.05\tablenotemark{2}
\\
 Hubble VIII     & 19 44 58.21  & -14 43 13.4  & 18.29 $\pm$ 0.03 & 1.24 & 0.36 &
0.80 & 0.44 & ~6.1 $\pm$ 0.3 & -6.24 & 0.93
\\
 \enddata
\tablenotetext{1}{Not available due to saturation at the cluster center.}
\tablenotetext{2}{Adopted from \citet{wyd00}.}

\end{deluxetable*}

\section{Observation}

Observations of NGC 6822 were made in Aug. 22 $\sim$ Sept. 23, 2003
(2003B) and in May 9 $\sim$ 17, 2005 (2005A)
using the MegaCam at CFHT in queue mode operation.
The MegaCam is a wide field mosaic camera composed of 36 2k $\times$ 4k
individual CCDs. It covers about $1 \deg \times 1 \deg$ area with
$0.187\arcsec$ per pixel resolution. Nine fields around the optical
main body of NGC 6822 were observed in ${\it g'}$ and ${\it i'}$ bands.
That makes our data coverage about 9 square degrees in total. Seeing was
on average 0.97 (0.90)$\arcsec$ in $g'$ band and 0.82 (0.75)$\arcsec$
in $i'$ band for the 2003B (2005A) season. The
variation of seeing condition between different seasons is negligible.
The integrated exposure time is 2100 sec (1980 sec) in $g'$ band and
1300 sec (1380 sec) in $i'$ band for the 2003B (2005A) season.
The raw images were processed using Elixir system by CFHT staff.
Elixir is a collection of programs, databases, and other
tools specialized in processing and evaluation of the large mosaic data.
Detailed information about the Elixir system can be found in \citet{mag04}.

\section{Discovery}

\begin{figure}
\plotone{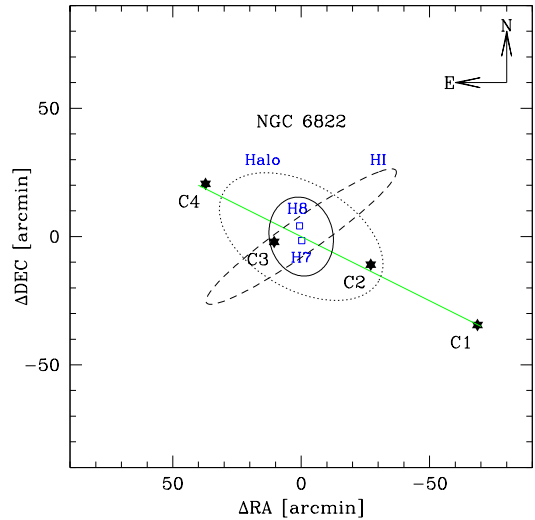}
\caption{Spatial distribution of star clusters in NGC 6822.
 The field of view is $3\degr \times 3\degr$ that was covered by
 our observation runs.
 The stellar symbols represent four new star clusters discovered in
our survey. All other star clusters of NGC 6822, including two
old clusters Hubble VII (H7) and Hubble VIII (H8) marked in squares are located in
the optical main body of NGC 6822 (a central ellipse in solid line).
An ellipse in dotted line indicates an approximate extent and shape
of NGC 6822 giant halo \citep{lee05,bat06} and an ellipse in dashed
line represents a schematic shape of HI disk \citep{deb00}.
Note that the new star clusters are spread widely, up to $77 \arcmin$
from NGC 6822 center for C1, and they are aligned in a line
that is almost coincident with the major axis of the giant halo and
perpendicular to the HI disk.}
\label{obs:map}
\end{figure}

We have visually inspected all image data and have discovered three new star
clusters (C1, C2, and C3) in the 2003B season data and another new star
cluster (C4) in the 2005A season data. The locations
of the four new star clusters are shown in Figure \ref{obs:map} and
the coordinates are listed in Table \ref{tab1}.
The first noteworthy point with these new star clusters is their
wide spatial distribution. The projected distances from the center of
NGC 6822 are $77 \arcmin$ for C1, $29 \arcmin$ for C2, $11 \arcmin$ for
C3, and $43 \arcmin$ for C4. Only one cluster C3 is located in the outskirts of
the ellipse defined by $15.5 \arcmin \times 13.5 \arcmin$, which represents
the approximate extent of the optical main body of this galaxy.
Previously known star clusters including H VII and H VIII \citep{KH04} are
inside this ellipse. The projected galactocentric distance of the most
distant cluster C1 reaches about 11 kpc in physical scale, which corresponds to
about 10 times the major axis radius of NGC 6822.

\begin{figure}
 \begin{center}
 \plotone{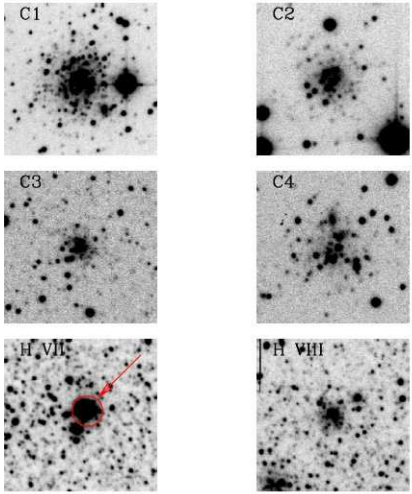}
 \caption{ Thumbnail images of four new star clusters (C1, C2, C3, and C4) compared
with NGC 6822 GCs, Hubble VII (marked by circle) and Hubble VIII in $i'$ band
observed using CFHT MegaCam.
The field of view of each image is $37\arcsec \times 37\arcsec$.
North is up and east is to the left.
Note that the new clusters are resolved into stars and more extended than
Hubble VII (H VII) which is not resolved in this ground based image.
}
\label{img}
 \end{center}
\end{figure}

Figure \ref{img} displays thumbnail $i'$ band images of these new star
clusters as well as H VII and H VIII in NGC 6822.
It is easily noted that the new
star clusters are more extended than H VII and are partially resolved into their
member stars, although the size and richness vary among clusters.
On the other hand, H VII, is not resolved into individual stars but
appears as a single extended source in the same image.
This indicates that the new star
clusters have different structural properties than typical globular clusters.
However, C3, the smallest new cluster, turns out to be similar to H VIII that is
another globular cluster with diffuse morphology and intermediate-age.

\section{Analysis and Results}
\subsection{Resolved Stellar Photometry}
\label{phot}

The analysis using the color magnitude diagrams (CMDs) based on
the resolved stellar photometry is one of the best ways to determine
the physical parameters of star clusters.
Since the new star clusters in NGC 6822
are partially resolved into individual stars in the current image data
(see Fig. \ref{img}), we construct CMDs of the stars to estimate the age
and the metallicity of the new ESCs.

We used DAOPHOT/ALLFRAME \citep{ste94} to carry out the photometry of
stars in the field where each star cluster was discovered.
Each field is $6.2 \arcmin \times 14.6 \arcmin$
defined by the size of one CCD chip among the 36 CCDs of MegaCam.
Standard procedures required by DAOPHOT/ALLFRAME such as initial
source detection, PSF selection, image montage, final source detection,
and final ALLFRAME photometry were followed.
For further information on this process, please refer to \citet{ste94}.
We adopted $4 \sigma$ threshold for the source detection.
The photometry data were calibrated using the standard transformation
relation provided by CFHT.

\begin{figure}
\plotone{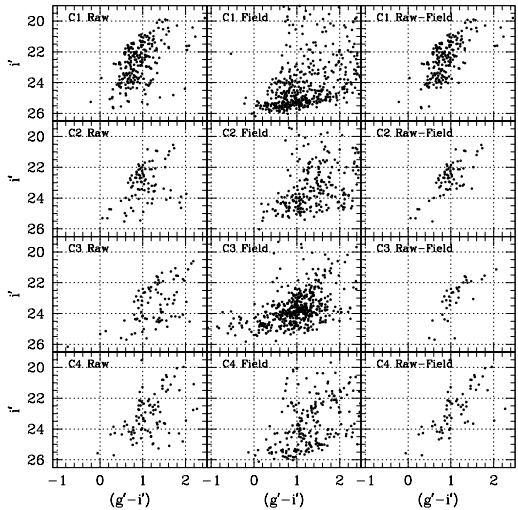}
\caption{
 The raw (left), field (center), and the field star subtracted (right)
 color magnitude diagram (CMD)s of the new star clusters, C1, C2, C3, and C4.
 See the text for details.}
\label{fscor}
\end{figure}

The number of detected sources in each field ranges from about
8000 (for C4 field) to 26000 (for C3 field).
We selected stars located within a certain concentric radius from each cluster
for the CMD analysis.
For C1, C2, and C4 the concentric radius of $15\arcsec$ was used,
while $10\arcsec$ is adopted for C3.
Among the selected stars, however, there are also many foreground stars,
which inevitably leads to heavy contamination to the target star clusters.
We corrected the field star contamination by using field stars collected
from the annulus set between $37\arcsec$ (200 pixels) and $74\arcsec$
(400 pixels) from each cluster center.
Figure \ref{fscor} shows the CMD of the new star clusters and their selected fields.
Firstly, we calculated the expected number of field stars for each CDM bin
with $0.5 \times 0.5$ mag by normalizing the number of field stars in the corresponding location of the field CMD.
Then, the field star correction was made by rejecting stars separated by less than $0.1$ mag from the expected field stars in each CMD bin of the cluster CMD.
The total number of removed field stars is 34, 25, 30, and 33,
leaving about 190, 79, 45 and 60 sources for C1, C2, C3, and C4,
respectively.
The rate of removed stars to all selected stars for each cluster,
ranging from 15\% (C1) to 40\% (C3), is consistent within the tolerance of $\pm 5\%$
even though we change the field star selection regions.
The field star subtracted CMDs of the new star clusters are displayed in
the right column of Figure \ref{fscor}.

\subsection{Reddening Estimation}
\label{reddening}

One difficulty in the interpretation of photometric results
of NGC 6822 clusters is that
NGC 6822 field suffers from a relatively heavy and patchy extinction
because it is located in low Galactic latitude
($b = -18.38926 \degr$). The foreground reddening estimated by
\citet{sch98} ranges from $E(B-V) = 0.16$ for the field of C1 to
$E(B-V) = 0.22$ for C2.
This requires the correction for patchy foreground extinction
for the analysis of the photometric properties of star clusters
that are widely spread in large area.

We utilized the observed colors of foreground stars, mostly Galactic
disk dwarf stars, to estimate the reddening for every region in our observed
field. The Galactic disk stars are observed as a vertical sequence
in $(g'-i')$ and $i'$ CMDs.
For a specific region, we select high $S/N$ and non-saturated stars
with $0.5<(g'-i')<1.5$ and $16<i'<19$ mag that lie within a circular area
of radius $3\arcmin$ and investigate their $(g'-i')$ color distribution.
The resultant color distribution is fitted by a Gaussian function to determine
the peak $(g'-i')$ color of the foreground stars and the approximate width of the
distribution.
However, if the number of high $S/N$ foreground stars is less than 300,
the size of circular area is increased by $0.1\arcmin$ to collect
more foreground stars for reliable determination.
The peak colors of selected foreground stars range from $(g'-i')=0.64$
to $1.16$ and the $(g'-i')$ color distributions are usually
very narrow with $FWHM \lesssim 0.05$ mag.

To convert this peak $(g'-i')$ color of the foreground stars to the
reddening value for the corresponding region, we used the reddening map by
\citet{sch98}
as a reference.
Firstly, we select about 220 locations evenly spaced over the $3\arcdeg \times
3\arcdeg$ field.
The separation between selected locations is about $12\arcmin$,
which is roughly twice the spatial resolution of reddening map by \citet{sch98}.
Then, we compare the peak $(g'-i')$ color of the foreground stars and the $E(B-V)$
estimate by \citet{sch98} for those selected locations to determine
the conversion factors that are required
to derive the reddening value for every location in the observed field.
More details on the reddening map construction around the field of NGC 6822
will be described in our forthcoming paper (Hwang et al. 2011, in preparation).

Finally determined and adopted reddening value for each cluster field
ranges from $E(g'-i')=0.16$ for C1 to $E(g'-i')=0.27$ for C2, as listed in
Table \ref{tab1}.
These reddening values are smaller than those estimated from \citet{sch98}
by about 0.1 mag in $E(g'-i')$, which is about 0.06 mag in $E(B-V)$.
We use these adopted reddening values for the analysis of CMD of the
new star clusters in NGC 6822.

\subsection{Color Magnitude Diagrams}
\label{cmdtxt}

\begin{figure*}
\plotone{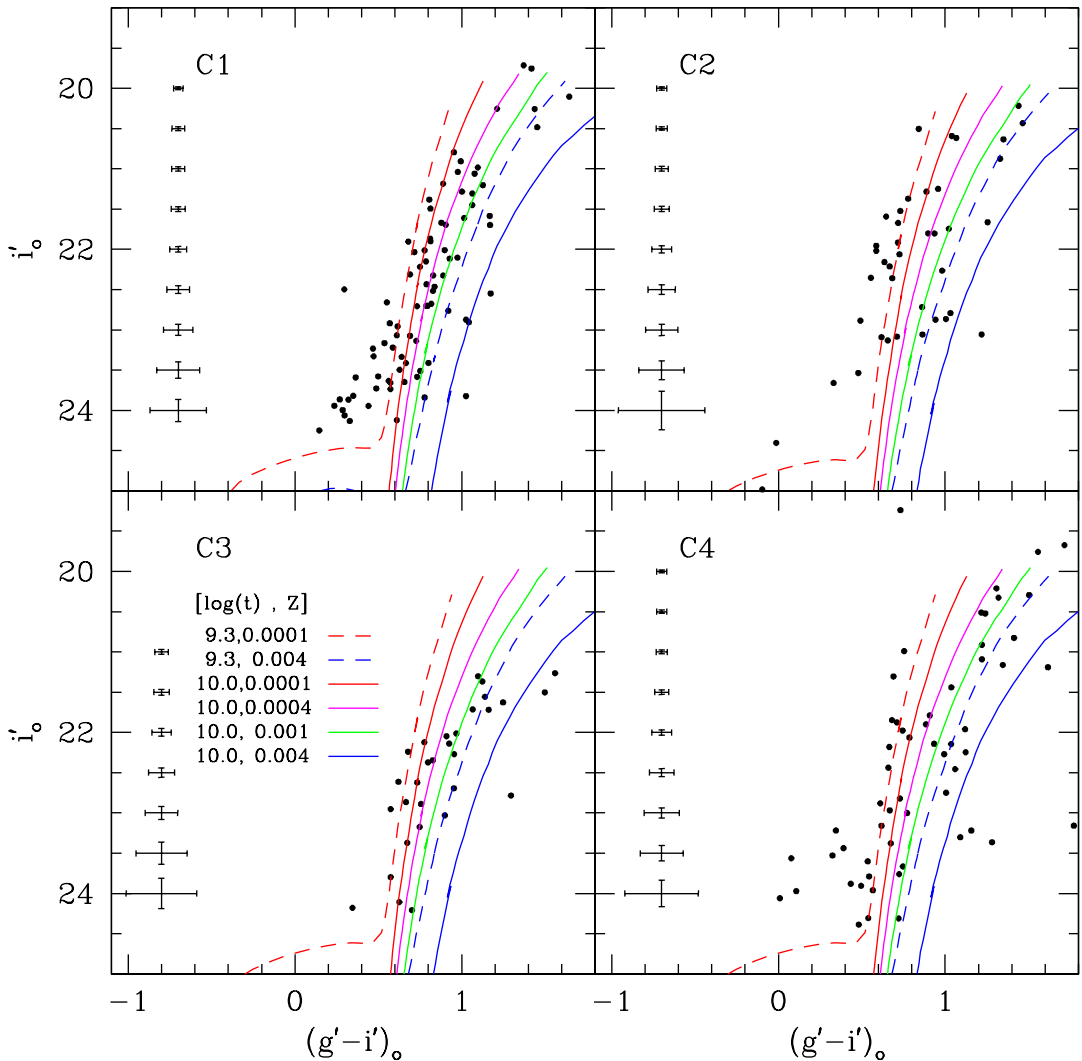}
\caption{ The $(g'-i')_{0}$ and $i'_{0}$ color magnitude diagram (CMD)s of
the new star clusters, C1 $\sim$ C4. Foreground stars were
statistically subtracted. Curved lines represent
the theoretical isochrones \citep{mar08}.
for Log(age)$=9.3$ (dashed lines) and 10.0 (solid lines), and
$Z=0.0001$, 0.0004, 0.001, and 0.004.
Each combination of
Log(age) and Z adopted for comparison is given in the lower-left
panel. See the text for details.} \label{cmd}
\end{figure*}

Figure \ref{cmd} shows the $(g' - i')_{0}$ -- $i'_{0}$ CMD of the resolved
stars in C1 $\sim$ C4. The adopted reddening for each cluster's field is
listed in Table \ref{tab1}.
The red giant branch (RGB) is clearly pronounced in each CMD, although the number of
detected RGB stars is different from cluster to cluster.
C1, the richest cluster among the four new clusters, exhibits a relatively well
populated RGB that runs from $i'_{0} \sim 20.0$ mag to $\sim 24.0$ mag.
The remaining three clusters also display similar RGB
features. However, there are still some notable differences.
The faint end of C2 RGB appears to be at $i'_{0} \approx 23.0$,
about one magnitude brighter than those of the other clusters.,
On the other hand, the bright end of C3 RGB is at $i'_{0} \approx 21.0$,
about one magnitude fainter than the other clusters.

We use the $i'$--band magnitude of the tip of RGB (TRGB) that is known to be constant
for the metallicity [Fe/H]$<-0.7$ \citep{lee93} for the estimation of the distance to
each cluster.
Even though it is hard to get rigorous measurement of the TRGB magnitude
due to the low number of sources, the CMDs in Figure \ref{cmd} indicate
that the TRGBs are located at $i'_{0} \approx 20.0$ for C1, C2, and C4.
Since the absolute magnitude of TRGB in $I$-band is $\approx -4.0$ mag
\citep{lee93},
which corresponds to $i' \approx -3.5$ mag,
the distance modulus for these three clusters is estimated to be $(m-M)_{0} \sim
23.5$.
This value is similar to the adopted distance modulus for NGC 6822,
$(m-M)_{0} = 23.35 \pm 0.02$ \citep{cle03,pie04,cio05}.
In the case of C3, though the fainter TRGB magnitude may suggest larger distance
than the other clusters, we assume that C3 is a member of NGC 6822 based on the
morphological similarity to H VIII as shown in Figure \ref{img}.
Therefore, we assume that the distance modulus for every cluster is the same as NGC
6822.

We have compared the observed cluster CMDs with the theoretical model
isochrones by \citet{mar08}  based on the adopted distance and
the foreground reddening for each cluster in Figure \ref{cmd}.
We selected two ages, Log$(t)=10.0$ (10 Gyr) and 9.3 (about 2 Gyr),
and four metallicity, Z=0.0001, 0.0004, 0.001, and 0.004
($-2.3 \la$ [Fe/H] $\la -0.7$).
The comparison with theoretical isochrones shows that RGBs of new star
clusters are broad so that isochrones with different metallicities and/or ages
can be overlaid at the same time. That is, the model isochrones with
Log$(t)=10.0$ and $0.0001 \le$ Z $\le 0.001$, and Log$(t)=9.3$ and
$0.0001 \le$ Z $\le 0.004$ can be used to represent the
majority of RGB stars in C1 and C2.
For C3 and C4, many bright RGB stars ($i'_{0}<22$) are traced by
the isochrones with $0.0004 \le$ Z $\le 0.004$.
This result suggests that the new clusters are as old as Log$(t) \gtrsim 9.3$
(about 2 Gyr) and their metallicities range from Z=0.0001 ([Fe/H]$\approx -2.3$)
to Z=0.004 ([Fe/H]$\approx -0.7$).

The photometric error in $(g'-i')$ is estimated to be no larger than 0.03 mag
when $i'_{0}<23.0$ mag, as marked in each panel in Figure \ref{cmd}.
Therefore, for $i'_{0}<23.0$ mag, the scatter in RGB may be real, suggesting
a metallicity dispersion. However, considering the crowded nature of clusters
and the limited spatial resolution of the current image data,
these broad RGBs could be the result of random errors involved in the source
detection.
Similarly, the study by \citet{hux05} show CMDs with broad RGBs for three ESCs
in M31 based on the ground-based imaging, overlaid with fiducial lines of
different metallicities ranging from [Fe/H]$=-2.17$ to [Fe/H]$=-0.47$. However,
in the subsequent study by \citet{mac06} based on the high resolution
Hubble Space Telescope (HST) image data, those M31 ESCs are confirmed to
have well-defined RGBs with no discernible metallicity dispersion.
Therefore, the future high resolution
observations are crucial for a better determination on the metallicity
and age of these new star clusters in NGC 6822.

\subsection{Size of Star Clusters}
\label{app}

\begin{figure}
\plotone{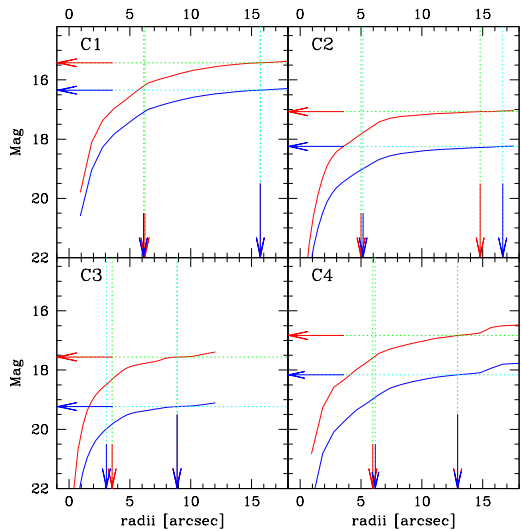}
\caption{
 Integrated magnitude profiles in $i'$ (red or thick line) and $g'$
 (blue or thin line) band of the new star clusters (C1, C2, C3, and C4).
Long and short arrows on $x$ axis mark
the total and half-light radii of each cluster, respectively, and arrows on $y$ axis
indicate the adopted integrated magnitude in $i'$ (red or thick line) and
$g'$ (blue or thin line) bands. Note that long arrows for C1, C3, and C4
and short arrows for C1 and C2 are overlapped.}
\label{prof}
\end{figure}

We have used APPHOT package in IRAF\footnote{IRAF is distributed by the
National Optical Astronomy Observatories, which are operated by the
Association of Universities for Research in Astronomy, Inc., under
cooperative agreement with the National Science Foundation.}
to derive the integrated luminosity profiles in $g'$ and $i'$ bands, and then to
determine the structural parameters of the new star clusters. For each star
cluster, we adopted incremental circular apertures with radii ranging from 4 to
100 pixels. The sky of each cluster was determined as the median of pixel
values in the 30-pixel-wide annulus set 150 pixel away from the cluster
center. We masked bright foreground stars to prevent any contamination to
the magnitude profile. The measured flux in each aperture was used to construct the
magnitude profile of the cluster as shown in Figure \ref{prof}.

In the case of an isolated star cluster, the integrated luminosity profile should
flatten out at the outer boundary of the cluster. However, due to the
field stars in the adjacent field, the integrated
luminosity profile may not converge to a fixed value but get brighter again at
certain radius, which is clearly seen in the profiles of C3, and C4.
Therefore, we determined the size of a cluster $R_{\rm t}$ by
finding the radius in the integrated magnitude profile where the magnitude
difference between adjacent annuli systematically decreases to minimum and,
at the same time, the magnitude difference between $R_{\rm t}$ and the
next outer annulus starts to increase again. We repeated this
procedure in $g'$ and $i'$ bands for each star cluster and determined the radii
$R_{\rm t}$ in both bands. At $R_{\rm t}$, the magnitude difference
between the adjacent annuli is usually about 0.01 mag or less. The cluster's
integrated magnitude was measured at the radius $R_{\rm t}$. The half light
radius $R_{\rm h}$ of a cluster was derived by interpolating along the
integrated magnitude profile to find a radius where the luminosity corresponds
to half or 0.75 mag fainter than the cluster's integrated magnitude.

Figure \ref{prof} shows the integrated magnitude profiles of the new star clusters
and
the determined values of $R_{\rm t}$ (long arrows) and $R_{\rm h}$ (short
arrows). The values of $R_{\rm t}$ and $R_{\rm h}$ in both $g'$ and $i'$
bands are usually in agreement with each other, differing by less than the
width of an annulus (2 pixels or $0.37\arcsec$). However, there are a couple
of cases that exhibit some differences in $R_{\rm t}$ and/or $R_{\rm h}$
depending on the filter bands. In the case of C2, the $R_{\rm t}$ is estimated
slightly larger in $g'$ band by about $1.85\arcsec$  than in $i'$ band, whereas
the $R_{\rm h}$ values in both bands turn out to be
consistent within the tolerance of $0.37\arcsec$. For C3, even with the same
$R_{\rm t}$ estimates in $g'$ and $i'$ bands, different $R_{\rm h}$ values
are derived, $3.07\arcsec$ in $g'$ and $3.54\arcsec$ in $i'$, which is the
largest discrepant measurement depending on the photometric bands.
It is regarded due to the difference in the observed profiles depending on
the filter band.
Therefore, we simply adopt the average of the estimates in both bands as final
values of $R_{\rm t}$ and $R_{\rm h}$ for the cluster.

The adopted values of $R_{\rm h}$ are $6.18 \pm 0.07 \arcsec$ ($14.0$ pc)
for C1, $5.09 \pm 0.08 \arcsec$ ($11.5$ pc) for C2,
$3.31 \pm 0.24 \arcsec$ ($7.5$ pc) for C3, and $6.08 \pm 0.11 \arcsec$ ($13.8$ pc)
for C4.
We also applied the same method to H VII and H VIII using our image data,
and derived $R_{\rm h} \approx 1.12\pm 0.04 \arcsec$ (2.5 pc) for H VII
and $R_{\rm h} \approx 2.67\pm 0.12 \arcsec$ (6.1 pc) for H VIII.
This indicates that
the new star clusters in NGC 6822 are larger than H VII, an analog of typical
old GC by a factor of $2 \sim 4$ and they are even larger than H VIII,
the diffuse GC in NGC 6822.
The adopted values of $R_{\rm h}$ and the associated errors are
listed in Table \ref{tab1}.

\subsection{Integrated Luminosity and Color}
\label{intphot}

We determined the total magnitude in $g'$ and $i'$ bands, and the $(g'-i')$
color of each star cluster based on the aperture photometry made with $R_{\rm t}$.
The $(g'-i')$ colors of new star clusters are relatively red, ranging from 0.93
(for C1) to 1.67 (for C3).\footnote{The $(g'-i')$ colors of the four new star
clusters were also measured
using smaller apertures with $r=3\arcsec$ and the result is $(g'-i')=0.91 \pm 0.01$
for C1, $1.22 \pm 0.01$ for C2, $1.50 \pm 0.03$ for C3, and $1.51 \pm 0.02$ for C4.
These values are
consistent with the colors determined using $R_{\rm t}$ and there appears to be
no systematic nor significant radial color gradient in the new star clusters.
Some differences in $(g'-i')$ colors are due to the selective inclusion or
exclusion of resolved bright RGB stars in the small aperture.}
The dereddened color
$(g'-i')_{0}$ is derived by correcting the adopted local reddening value
$E(g'-i')$ for each star cluster, determined in Section \ref{reddening}, which is
$(g'-i')_{0}=0.77$, $0.90$, $1.44$, and $1.16$ for C1, C2, C3, and C4, respectively.
For comparison with other studies, we have converted the $g'_{0}$ and
$(g'-i')_{0}$ of each star cluster into $V_{0}$ and $(V-I)_{0}$ using a
transformation equation derived by comparing the same Landolt standard
stars observed in $g'$ and $i'$ bands \citep{smi02} with those observed
in $V$ and $I$ bands \citep{lan92}: $V_{0} = g'_{0} - 0.370 \times (g'-i')_{0} -
0.061$
(RMS = 0.015), $(V-I)_{0} = 0.699 \times (g'-i')_{0} + 0.312$ (RMS = 0.014).
The absolute magnitude
$M_{V,0}$ of each cluster ranges from $M_{V,0} = -7.70$ for C1 to
$-5.22$ for C3. The $(V-I)_{0}$ colors of the new clusters range from
$(V-I)_{0} = 0.85$ for C1 to $1.31$ for C3.
The adopted integrated magnitude and color of new star clusters as well as
the reddening values are listed in Table \ref{tab1}.

The luminosity and color of H VII are
$M_{V} = -7.54$ and $(V-I) = 1.31$ \citep{wyd00}
\footnote{Because H VII has its central part saturated,
we adopt the photometric data from the study by \citet{wyd00}.}.
Adopting $E(B-V)=0.19$ that was derived using the color of foreground stars
in the H VII field, the reddening corrected magnitude and color of H VII are
$M_{V,0} = -8.16$ and $(V-I)_{0} = 1.05$. For H VIII,
we have derived the total magnitude $g'=18.29 \pm 0.03$ and
color $(g'-i')= 1.24 \pm 0.03$ using our image data and have determined
the reddening $E(g'-i')=0.36$ or $E(B-V)=0.21$.
Finally calibrated absolute magnitude and
color of H VIII are $M_{V,0} = -6.24$ and $(V-I)_0 = 0.93$. This shows that
the magnitudes and colors of these GCs in NGC 6822 are similar to those
of the Galactic GCs.

The fact that the $(V-I)_{0}$ colors of the new star clusters are similar to
or redder than those of H VII and H VIII in NGC 6822, suggests
that the new clusters are as old as these two GCs.
We have compared the $(V-I)$ color expected by the theoretical model \citep{mar08}
with the integrated $(V-I)_{0}$ colors of the new star clusters in NGC 6822.
For Z=0.0001 ([Fe/H] $\approx -2.3$), similar to the metallicity
[Fe/H]$\approx -2.0$ of H VII \citep{coh98}, the model predicts $(V-I) \ga 0.80$ for
Log$(t)>10.0$ so that every new star cluster should be older than Log$(t)=10.0$.
This expects C3 and C4 to be even older than Log$(t)=10.1$,
the maximum age (about 12.5 Gyr) provided by the model,
which requires higher metallicity for these two clusters.
When we assume Z=0.004 ([Fe/H] $\approx -0.7$), the metallicity of HII regions in
NGC 6822 \citep{skil89},
the ages of C1 and C2 are consistent with $9.0<$Log$(t)<9.5$, while
the estimated age of C3 and C4 is still Log$(t) \geq 10.0$.

This is basically in agreement with the results obtained in Section \ref{cmdtxt}
that the new clusters are older than Log$(t)=9.3$ and relatively metal poor with
$0.0001 \le$ Z $\le 0.004$. However, the extremely red integrated colors
of C3 and C4 may imply higher metallicity compared to the other clusters or
the higher reddening for the corresponding field.
The CMD of C3 in Figure \ref{cmd} shows that the bright end of RGB is at about
$i'_{0} = 21.0$, one magnitude fainter than those of the other clusters.
If we assume that the faint RGB is due to the intrinsic or the unaccounted
reddening and the unreddened tip of RGB (TRGB) should be around $i'_{0} = 20.0$,
then the extra reddening of $E(g'-i') \approx 0.4 - 0.5$ has to be introduced,
making the total reddening of the C3 field $E(g'-i') \approx 0.63-0.73$.
This would make the reddening corrected color of C3 $(V-I)_{0} \approx 0.97-1.04$,
similar to the color of H VII and H VIII.
However, if the reddening for C3 field is greater than $E(g'-i') \approx 0.44$,
the theoretical isochrones cannot be fitted to the CMD of C3 in Figure \ref{cmd}.

If we accept that the red colors are intrinsic properties of the star clusters,
then it would be worthwhile to note that the two new clusters C3 and C4 are as red
as $(V-I)_{0}> 1.0$ and relatively extended with $R_{\rm h}>7.0$ pc,
which satisfies the criteria for faint fuzzy clusters.
This is the first sample of faint fuzzy clusters discovered
in the Local Group. We will discuss this in Section \ref{ecsetc}.

\begin{figure}
 \plotone{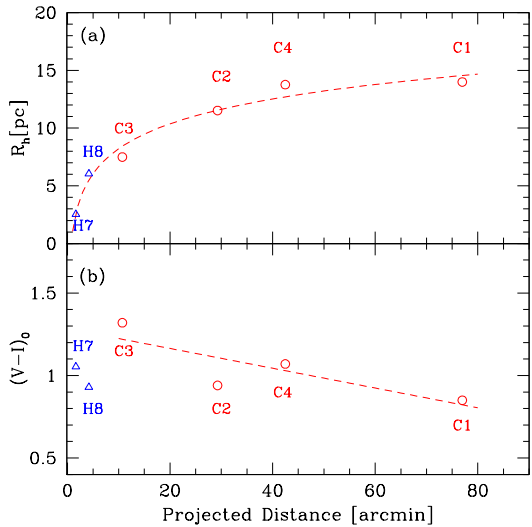}
 \caption{ The half-light radii $R_{\rm h}$ (panel a) and
$(V-I)_{0}$ color (panel b) distribution of the new star clusters
as well as H VII and H VIII along the projected galactocentric
distance of NGC 6822. The dashed line in panel (b) is a simple
linear fit to the radial distribution of $(V-I)_{0}$ colors of
the four new clusters.}
\label{rdist}
\end{figure}

\section{Discussion}

\subsection{New Star Clusters and Old Stellar Halo}
\label{oldpop}

The wide spatial distribution of newly discovered star clusters
is quite surprising considering
that the host galaxy NGC 6822 is a small dwarf irregular galaxy.
The new star clusters are spread in the area as large as about
$120\arcmin \times 80\arcmin$, while the existing star clusters
are concentrated in the optically visible main body of NGC 6822 that is
about $15.5 \arcmin \times 13.5 \arcmin$ wide at most.
This is even much larger than the area spanned by old and intermediate
stellar halo, which extends out to about $40\arcmin$ \citep{lee05,bat06}.
This is similar to the disks of satellite galaxies in the Milky Way Galaxy
\citep{met07}.

There are a couple of noteworthy points regarding
the new star clusters and the old stellar halo of NGC 6822.
Figure \ref{obs:map} shows that the four star clusters C1 -- C4 are
distributed along the projected line running from
northeast (NE) to southwest (SW). This is consistent
with the major axis of old stellar halo, which is elongated in the
same direction (NE-to-SW) as marked in dotted lines in Figure \ref{obs:map}.
The other point is that the four new star clusters are expected
to be older than Log$(t) \approx 9.0$ based on the CMD and integrated
color analysis.
The RGB stars that are used to trace the old halo of NGC 6822 are
also relatively old population with Log$(t)\ga 9.0$.
These suggest that the old stellar halo and the new old star clusters
may share the common formation history.

Figure \ref{rdist} displays the distribution of half-light radii $R_{\rm h}$
and $(V-I)_{0}$ colors of the four new clusters
as well as H VII and H VIII along the projected distance from the NGC 6822 center.
Interestingly, as the clusters are located farther from the galaxy center,
the clusters tend to be larger in size and bluer in color.
The change of the cluster sizes along the distance can be fitted by
a log-square function as shown in dashed lines in Fig.\ref{rdist}(a).
If we consider only the four new clusters, there also seems to be a systematic
color change with its gradient about $-0.006 \pm 0.003$ mag/arcmin, marked in
dashed lines in Fig.\ref{rdist}(b).
If the red color of C3, however, is mainly due to the unaccounted intrinsic
reddening as discussed in Section \ref{intphot}, the color gradient along
the projected distance would be not so significant.

A similar kind of size distribution is also pointed out for Milky Way and LMC GCs
in the study of \citet{vdb04}.
In the outer Galactic halo with galactocentric distance D$_{gc}>40$ kpc,
only extended GCs with $R_{\rm h}>15$ pc are found, while the inner halo
is populated with both compact and extended GCs.
Similarly, LMC GCs with $R_{\rm h}>10$ pc are only found in the region
farther than $8\degr$ from the galaxy center.
This dependence of star cluster size
on the galactocentric distance may be interpreted as the result of
high survival rate of extended clusters from the disruption in the
outer halo where relatively weak tidal forces are exerted upon the clusters
by the galaxy itself.
The similar trend observed in the size distribution of star clusters
in Figure \ref{rdist} shows that the old cluster system of NGC 6822 has
a common characteristic with Galactic GC as well as LMC GC systems.

Another noteworthy point is that the extended GCs with $R_{\rm h}>10$ pc
are usually metal poor with [Fe/H]$\la -1.3$, as shown in Figure 5 of \citet{vdb04}.
The metallicity of the four new clusters is estimated to be
$0.0001 \le$ Z $\le 0.001$ ($-2.3 \le$ [Fe/H] $\le -1.3$) if we assume
the age of clusters Log$(t)=10.0$.
If we interpret the color distribution in Figure \ref{rdist}(b)
as metallicity distribution, it leads to strong correlation between the
metallicity and the cluster size, that is, the lower metallicity for
the larger clusters. This would be consistent with
the metallicity distribution of extended Galactic GCs.
This may also suggest the metallicity gradient in the halo
of NGC 6822, lower metallicity in the outer halo than in the inner halo.

 \begin{center}
 \begin{figure*}
 \includegraphics[scale=0.6,angle=-90]{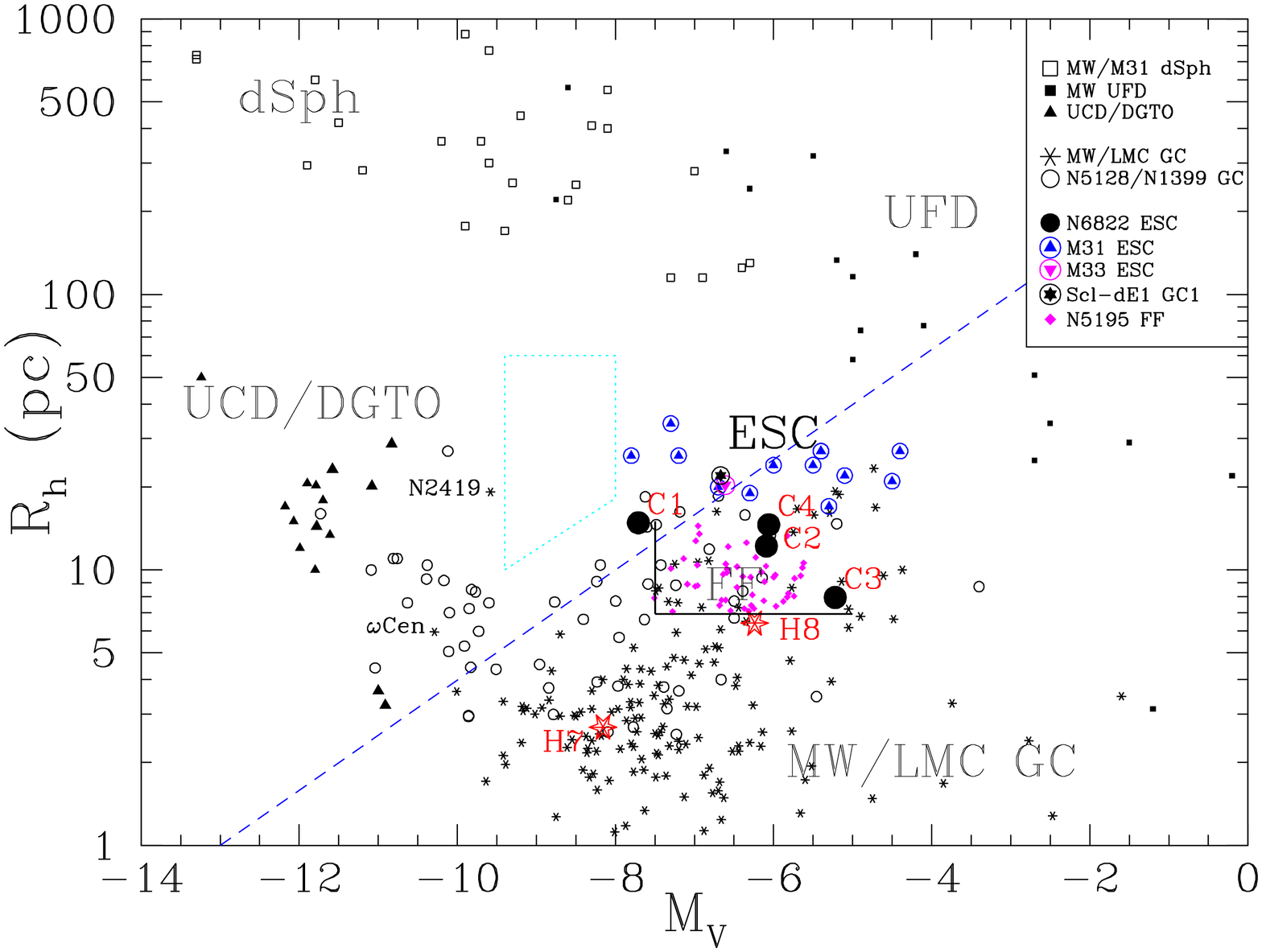}
 \caption{
$M_V$ vs. half light radii ($R_h$) diagram of GCs including the
Galactic GCs and LMC GCs \citep{vdb96,vdb04}, ESCs including M31 ESCs
\citep{hux05,hux11} and M33 ESC \citep{sto08} as well as NGC 6822 ESCs.
Two Hubble clusters in NGC 6822, Hubble VII (H7) and Hubble VIII (H8) are
also marked for reference.
Dwarf galaxies are satellites of the Milky Way Galaxy\citep{irw95,kal10} and M31
\citep{mcc06, mcc08, mar06, mar09, irw08, iba07, maj07, col10, kal10}.
Overplotted data include NGC 5128 GCs \citep{gom06,mar04,mou10},
NGC 1399 GCs \citep{ric05}, UCDs \citep{dep05,mie02} and DGTOs \citep{has05},
 along with the ultra faint dwarf (UFD) galaxies
recently discovered in the Milky Way Galaxy \citep{mar08b, dej10, bel09, bel10}.
The dashed line is the locus Log$(R_{\rm h}) = 0.2 M_{\rm V} + 2.6$
set by \citet{vdb04} as the upper limit for typical GCs.
The L-shape indicates the region where faint fuzzy clusters are found
as shown by those in NGC 5195 \citep{hwa06}.
Note that C1 lies above the GC locus as are the ESCs in M31 and M33,
while C2, C3, and C4 are located in the L-shape region, below the GC locus.
While traditional Galactic dSphs and M31 dSphs
are clearly separated from typical GCs in this diagram,
many luminous and compact objects such as UCDs and DGTOs as well as faint
objects such as ESCs and UFDs are found to be mixed with GC populations.
A box in dotted line represents an `avoidance zone' defined by
$-9.4 \lesssim M_{V} \lesssim -8.0$ and $10 \lesssim R_{h,pc} \lesssim 60$.
See the text for details.}
\label{radist}
\end{figure*}
\end{center}

\subsection{Complex of Old and Young Stellar Structures}
\label{complex}

The spatial distribution of old stars combined with
old extended clusters makes stark contrast to the central part of
NGC 6822 dominated by young stellar populations.
This shows that a small dwarf irregular galaxy NGC 6822 is
actually a complex of two distinct stellar structures, old stellar halo
and young central body.
Recent studies also show that the young stellar populations are not
restricted to the central body of NGC 6822 but are distributed
around the large HI gas structure \citep{komi03, deb03} extending
northwest-to-southeast direction \citep{deb00}.
This HI gas structure with blue stars with ages of $\sim 10^8$ yr
is almost perpendicular to the alignment of the new star clusters and
the elongated old stellar halo (see Figure \ref{obs:map}).

NGC 6822 is not the only one dwarf irregular galaxy known to have such
a complex stellar structure.
The co-existence of underlying old stellar population and the gas-rich
young stellar structure is also reported in other dwarf irregular galaxies,
including Leo A \citep{van04}, LMC \citep{min03}, and IC 10 \citep{dem04}.
The origin of such complex stellar structures in dwarf irregular galaxies
is not clearly understood yet.
However, it has been recently suggested that the merging between gas-rich
dwarf galaxies may induce the formation of the outer old and the inner
young stellar structures \citep{bek08}.

Even though it is relatively common to have old underlying stellar
structure in dwarf irregular galaxies, ESCs associated
with the old stellar halo are not discovered in other dwarf galaxies
but only in NGC 6822.
If the old stellar structure is the remnant of the merged galaxy
as suggested by some theoretical studies, the old ESCs
should have survived the merging or accretion process,
which is an important clue to the understanding of their origin.

\subsection{Nature of ESCs}
\label{ecsetc}

ESCs in small dwarf galaxies are rarely known.
NGC 6822 and Scl-dE1, a dwarf spheroidal galaxy in the Sculptor group
\citep{dac09}, are the only cases where ESCs are reported to date.
However, if we define the ESCs only by their large size, then such extended
clusters may not be totally new. Even in the Milky Way galaxy, there are
already about 10 GCs with R$_{h}>10.0$ pc and the largest GC, Palomar 14,
is even as large as R$_{h} = 24.7$ pc \citep{vdb96}, which could be easily
classified as ESCs.
Nevertheless, even those extended GCs are found to satisfy a certain
criterion set between the luminosity and the size of clusters, that is,
log$R_{h} = 0.2 M_{V} + 2.6$ outlining the upper limit locus in $R_{h}$
versus $M_{V}$ parameter space \citep{vdb04}.
Figure \ref{radist} shows that GCs in LMC, as well as the Galactic GCs,
also satisfy the same criterion.

On the other hand, the exceptionally bright and large GCs, NGC 2419 and
$\omega$Cen are found not to satisfy the above criterion. Instead, these two
GCs appear to occupy the similar parameter space with very small dwarf
galaxies such as ultra compact dwarf (UCD) galaxies \citep{mie02}, as shown in
Figure \ref{radist}. Moreover, $\omega$Cen is known to have multiple stellar
populations and is regarded as the core of a disrupted dwarf galaxy \citep{lee99}.
These results have led to consider that the criterion set by \citet{vdb04}
could be used to separate typical GCs from other types of compact objects
such as core of dwarf galaxies.

However, recently numerous observational studies have led to discoveries of star
clusters that do not satisfy the criterion set by \citet{vdb04}. Some examples
include GCs in a giant elliptical galaxy NGC 5128.
\citet{gom06} show that new GCs in NGC 5128 fill in the gap that used to
exist around the GC locus set by \citet{vdb04} in the parameter space of
$R_{h}$ versus $M_{V}$ over a large range of luminosity ($-11<M_{V}<-5$).
Especially, combined with newly discovered luminous GCs in NGC 1399 \citep{ric05},
Figure \ref{radist} shows that the new NGC 5128 GCs appear to smoothly
connect typical GCs with the existing luminous GCs including NGC 2419 and
$\omega$Cen, crossing the GC locus. The
major characteristics of these newly discovered clusters are the high
luminosity and the extended structure, larger than typical GCs. Since these
properties are also common to UCDs and `dwarf-globular transition objects
(DGTO)' in the Virgo cluster \citep{has05}, some of the luminous GCs may be
regarded as remnant or stripped core of dwarf galaxies, such as in the case of
$\omega$Cen.

We have added the four ESCs and two GCs (H VII and H VIII)
in NGC 6822 as well as 13 ESCs in M31 \citep{hux05,hux11} and
one ESC in M33 \citep{sto08} to the $R_{h}$ versus $M_{V}$ plot in
Figure \ref{radist}. It is clear that four M31 ESCs, M33 ESC, and C1
in NGC 6822 lie above the locus that defines the upper limit for typical
GCs, while the other new clusters C2, C3, and C4 in NGC 6822 as well as
the other ESCs in M31 occupy the same parameter space with typical GCs.
With the addition of GCs in NGC 5128, the distribution of star clusters
with $-8<M_{V}<-5$ over the GC locus set by \citet{vdb04} appear rather
smooth, implying that ESCs could be an extension or a family of typical
GC population.
It is also noted that every sample of ESCs discovered to date usually has almost
the same magnitude range covering $-8.0 \lesssim M_{V} \lesssim -4.5$.

However, ESCs in M31, M33, and NGC 6822 are quite different from
the luminous GCs in NGC 5128 and NGC 1399 since the
ESCs are more than 2 magnitude fainter in $V$ band.
Between these two classes of clusters, there appears to be a zone defined by
$-9.4 \lesssim M_{V} \lesssim -8.0$ and $10 \lesssim R_{h,pc} \lesssim 60$,
marked by a dotted box in Figure \ref{radist}, where few star clusters
and/or dwarf galaxies are found.
Even NGC 5128 GCs that span the large magnitude range from $M_{V}=-10$
to $-5$, do not occupy this `avoidance zone'. This leads to a speculation that
faint ESCs may be physically different from the luminous
extended GCs. We will discuss more about this in Section \ref{ori}.
However, further observational investigations are required over
other GC systems to confirm whether the `avoidance zone' is physically real.

Another class of star clusters that is worth consideration is the faint fuzzy
clusters. Faint fuzzy clusters have large size (R$_{\rm h} >7$ pc) and faint
luminosity ($M_{V} > -7.5$) as marked by two rectangle lines in Figure
\ref{radist}. One more characteristic of faint fuzzy clusters is the red color
($V-I>1.0$), slightly redder than typical GCs.
These clusters were first introduced by \citet{lar00}
and are known to exist only in some nearby SB0 galaxies including NGC 1023,
NGC 3384, and NGC 5195 \citep{lar01,hwa06}.
It has been suggested that faint fuzzy clusters are relatively old
($\ge 7 - 8$ Gyr), as massive as $10^4 - 10^5 M_{\odot}$ \citep{lar02,hwa06},
and moderately metal rich with [Fe/H]$\sim -0.6$ \citep{lar02}.
From Figure \ref{radist}, it is noted that faint fuzzy clusters satisfy
the criterion set by \citet{vdb04} as shown by the sample of faint fuzzy
clusters in NGC 5195 \citep{hwa06}, suggesting that the faint fuzzy clusters
could be another population of old GCs.

We have shown that, among the four new star clusters in NGC 6822, two
clusters C3 and C4 can be classified as faint fuzzy clusters with $(V-I)_{0}
\gtrsim 1.0$.
The other two clusters C1 and C2 have $(V-I)_{0}$ colors of typical GCs,
which excludes them from the class of faint fuzzy clusters. The
comparison with theoretical isochrones as well as model integrated colors implies
that
the metallicity of C3 and C4 may be Z$\ga 0.004$, i.e., [Fe/H]$\ga -0.7$.
This is consistent with the general picture of faint fuzzy clusters with
moderately higher metallicity than that of typical GCs \citep{bro02}.
These results suggest that the two new extended star clusters, C3 and C4,
may be the first sample of faint fuzzy clusters in the Local Group.

The faint fuzzy clusters have been discovered only in three SB0 type galaxies
including NGC 1023, NGC 3384, and NGC 5195 and these clusters are
regarded as disk population, following disk rotation confirmed in NGC 1023
\citep{bro02}. Therefore, discovery of faint fuzzy clusters in NGC 6822 is
very interesting in the following two points: (1) NGC 6822 is not an SB0 type
but dwarf irregular galaxy. This makes the first case of faint fuzzy clusters
discovered in a galaxy that is not SB0 type. (2) The two extended clusters C3
and C4 are located in the halo, not in the disk-like structure, of NGC 6822.

A noteworthy point regarding the origin of faint fuzzy clusters is that the
existing faint fuzzy clusters are found in galaxies that have undergone
dynamical interactions with neighboring galaxies. Therefore, the formation of
these peculiar clusters is suspected to have correlation with the dynamical
interactions of host galaxies (e.g. \citealt{fel02}). As mentioned in Section
\ref{complex}, NGC 6822 is a complex of old and young stellar populations
superimposed upon a giant HI disk-like structure and such a complex structure
is suspected to be the product of merger-like events that may have triggered the
very recent increase of star formation rate (e.g., \citealt{gal96c}).
Therefore, the existence of faint fuzzy clusters and ESCs may be
consistent with the complex structure of NGC 6822.

\subsection{Origin of Extended Star Clusters}
\label{ori}

As many ESCs are discovered in the very outer part of
galaxies, their origin and formation mechanism
are of great importance for the understanding of the galaxy halo as well
as ESCs themselves. Currently, there are several views on the origin or
the formation of ESCs, which can be roughly categorized into the following
three scenarios:
\begin{enumerate}
 \item The remnant or core of tidally stripped dwarf galaxies form ESCs.
 \item Collisions of two or more star clusters in star cluster complexes or
     super-star cluster make ESCs.
 \item Star clusters are born in various sizes and some ESCs survive the
     disruption under the weak tidal field.
\end{enumerate}
Without spectroscopic and deeper imaging observations at the current stage,
it is difficult to know the origin of ESCs in NGC 6822. However, we
briefly discuss the above three scenarios one by one in the context of dwarf
irregular galaxy NGC 6822.

\subsubsection{Remnant of Dwarf Galaxies}

Since the size of many ESCs are of similar to those of
very small dwarf galaxies such as UCDs, the
ESCs have been suspected to be remaining cores of tidally disrupted dwarf
galaxies. A representative example is $\omega$Cen in the Milky Way \citep{lee99}.
If these ESCs are remaining cores or remnants of dwarf galaxies, then one of
some important characteristics would be their stellar contents of multiple
populations. The study by \citet{mac06} shows that the four ESCs in M31,
observed with HST, do not reveal any signature of multiple stellar populations
but only display features of typical old GCs in the CMD analysis.
The spectroscopic observation of resolved stars in one M31 ESC by
\citet{col09} also suggests that the cluster is not a dark-matter dominated system.
These results of observational studies are against the dwarf galaxy core scenario for
the origin of ESCs.

However, one important difference between clusters like $\omega$Cen and
M31 ESCs is the luminosity. As shown in Figure \ref{radist},
many luminous GCs with $M_{\rm V} \la -9.5$ share the same parameter
space with $\omega$Cen and UCDs. On the other hand, M31 ESCs are
$M_{\rm V} > -8.0$, fainter than  $\omega$Cen-like clusters by more than
2 magnitude. These two classes of ESCs are separated in $R_{\rm h}$
and $M_{\rm V}$ space by the so-called `avoidance zone' defined in
Section \ref{ecsetc} and marked in Figure \ref{radist}.
Therefore, these two classes may be physically different populations.

Like M31 ESCs, NGC 6822 ESCs are also fainter than $M_{\rm V} = -8.0$,
located on the fainter side of the `avoidance zone' in Figure \ref{radist}.
Therefore, it is likely that NGC 6822 ESCs should be similar to
M31 ESCs in many respects, rather than the remnant or the core of defunct
dwarf galaxies. However, deep imaging and
high resolution spectroscopic observations are required to test this possibility.

\subsubsection{Star Cluster Collisions}

The mechanism of star cluster collisions in cluster complexes or
super star clusters to form ESCs has been suggested by \citet{fel02}
to explain the origin of faint fuzzy clusters in NGC 1023.
This scenario requires the interaction of galaxies to induce strong star formation
and to form super star clusters or stellar complexes within which collisions of
individual clusters would lead to form eventually ESCs or faint fuzzy clusters.
Actually, some super star clusters or star cluster
complexes have been reported in galaxies that are undergoing dynamical
interactions, such as the Antennae galaxy \citep{whi99,whi10} and M51 \citep{bas05}.

The hierarchical structure of young stellar population has been recently
reported in the central part of NGC 6822 \citep{kar09, gou10},
although the estimated mass of these stellar groupings or
associations is relatively small ($10^3 - 10^4 M_{\odot}$) compared to
super star clusters in other galaxies
(e.g., $\ga 10^5 M_{\odot}$ in the Antennae).
It may still provide some chance of star cluster collisions within the
hierarchical structure at the central part of NGC 6822 due to the high spatial
density.
Since NGC 6822 ESCs have similar old age and are spread over the wide area where
no specific structure is observed other than rather smoothly distributed old stars,
formation of these ESCs from several regions of high stellar density
would require strong star formation events over a huge volume enveloping
the whole stellar halo of NGC 6822.
This may not be consistent with the star formation history of NGC 6822 that
strong star formation has only been initiated very recently, about $\sim 10^8$ yr ago
\citep{gal96c}. Star formations are also expected to begin from the very early
stage of stellar evolution of NGC 6822, possibly $\sim 10$ Gyr ago \citep{gal96b},
but it is not clear whether the old star formation was strong enough to form
large stellar complexes.

In a very recent study by \citet{ass11},
it has been proposed that the merger of two star clusters
within low mass dark matter haloes could produce an ESC. It
involves very small mass of dark matter halo so that the velocity
dispersion at the cluster center would be $0.7 - 1.7$ km sec$^-1$,
which is beyond the measurement accuracy of the current facilities.
The velocity dispersion of one ESC, EC4, in M31 is estimated to be
$2.7^{+4.2}_{-2.7}$ km sec$^{-1}$ \citep{col09}.
Therefore, according to this theory, the ESC could still have a small mass
dark-matter that help to sustain the extended structure.

\subsubsection{Intrinsic ESCs -- Survivors of Tidal Disruption}

Another recent theoretical study by \citet{hur10}
argues that star clusters can be born very extended and the clusters could
evolve to become old ESCs if the clusters are placed under the weak tidal
field, which would be possible only in the outskirts
of dwarf galaxies such as NGC 6822.
This also suggests that, under the strong tidal field, it would be impossible
for star clusters to survive to become old ESCs.
This implies that ESCs, even those observed in a large spiral galaxy like M31,
are supposed to form in small dwarf galaxies.

Based on a wide-field survey of M31 halo, \citet{mac10b}
show that many halo GCs of M31 are likely to be found close to stellar streams
and that about one-third of the newly discovered halo GCs are of the
extended nature. Therefore, it is suggested that M31 ESCs have
assembled into M31 halo as a consequence of accretion from the proto-dwarf
galaxies that should have disrupted leaving stellar streams behind.
This is also consistent with the observational result
that the spectroscopically measured velocities of one ESC EC4 in
M31 and a nearby stellar stream are in agreement with each other \citep{col09},
indicating the the ESC and the stellar stream are physically associated.

The systematic variation in the sizes of ESCs along the galactocentric
distance of NGC 6822, as shown in Figure \ref{rdist}, may be consistent with
this scenario in terms that the smaller clusters (e.g., C3) are located closer
to the galaxy center where the tidal interaction would be strong,
while the larger cluster (e.g., C1 and C4) in the outer halo where the tidal force
would be weak.
Combined with the systematic change of color along the galactocentric distance,
it may support the idea that the ESCs have formed as part of the old stellar halo of
NGC 6822.
This is also consistent with the suggestion that ESCs may be
an extension or another family of typical GCs.

However, this poses an intriguing question about why NGC 6822 was more
effective to form ESCs over the other dwarf galaxies in the Local Group.
There are not many dwarf galaxies that possess old GC system. Only about four
dwarf irregulars including LMC and about five dwarf spheroidals among more
than 30 dwarf galaxies in the Local Group are known to have their
own GC system \citep{vdb00}.
There is even no known dwarf galaxies in the Local Group
that have any ESC other than NGC 6822.
Unlike NGC 6822 that is relatively isolated without neighbors, many other
dwarf galaxies are found in groups concentrated around two giant spiral galaxies,
the Milky Way and M31, in the Local Group.
The environmental condition of NGC 6822 might have contributed to create
the optimal tidal field to have ESCs.
However, further wide and deep observations around other dwarf galaxies
are required to find any answer to this question.

\subsubsection{ESCs -- Mixture of Heterogeneous Populations}

If we define ESCs as GCs with large physical size,
$R_{\rm h} \ga 7$ -- $10$ pc, then the origin of ESCs observed
in many galaxies seems to be diverse depending on the environment and
the evolutionary history followed by the host galaxies.
At least, there may be two populations of ESCs that have different
formation mechanisms.

The first class includes ESCs that are relatively luminous with $M_{\rm V}<-9.5$
mag and are located in the brighter side of the `avoidance zone'
in the $M_{\rm V}$ and $R_{\rm h}$ parameter space in Figure \ref{radist}.
Examples include a few Galactic GCs such as
$\omega$Cen and NGC 2419, some luminous GCs in NGC 5128 and NGC 1399.
Since these ESCs are comparable to UCDs and DGTOs in terms of the luminosity
and size, it would be reasonable to assume that these luminous ESCs and
compact galaxies should have common origins. The fact that some of
these ESCs are known to have signs of multiple stellar populations
(e.g., $\omega$Cen) can be supportive of the argument that ESCs are
the cores of tidally disrupted dwarf galaxies.
It would be also very interesting to test whether some of these luminous
ESCs could be remnants of super star clusters or large stellar complexes.

The second class of ESCs is composed of relatively faint clusters
including extended clusters in M31 and M33 as well as faint fuzzy clusters.
These ESCs are less luminous and
are located in the fainter side of the `avoidance zone' in Figure \ref{radist}.
The ESCs discovered in NGC 6822 belong to this class.
These ESCs may share the same formation mechanism with typical GCs
but have evolved under the specifically optimized tidal field.
Faint fuzzy clusters may be a sub-population of this class with slightly
enriched metallicity.
Dwarf galaxies would be the preferred place
for the formation of such ESCs due to the relative weak tidal forces exerted
by less massive galaxies, as suggested by \citet{hur10}.
However, dwarf galaxies with ESCs seem to be rare.
Only two dwarf galaxies are known to date to have ESCs:
NGC 6822 in the Local Group and Scl-dE1 in Sculptor Group \citep{dac09}.
Future observations over the large number of dwarf galaxies would be
valuable resources to test this idea on the origin of ESCs.

These two classes of ESCs, regardless of their luminosities, seem to be
intermediary objects between typical compact GCs and large dwarf galaxies.
A similar idea has been proposed by \citet{nor11} on the study of UCDs
arguing that UCDs may be composed of more than two physically distinct
populations including stripped nuclei of galaxies, giant GCs or super star clusters,
and disrupting GCs based on a tight correlation between mass and size of those
objects.
Adopting this scheme, the first class ESCs belong to giant GCs or stripped nuclei
of galaxies, while the second class ESCs belong to the disrupting GCs.
Between these two different classes, there exists a region where neither the first
class nor the second class are found (see their Figure 16), which corresponds
to the `avoidance zone' in Figure \ref{radist}.

One more point that might be worth noting is that there is still a
possibility that some ESCs found in the outermost halo of galaxies
might not be bound to the gravitational potential of the host galaxies.
A recent report on the intergalactic GCs in the Virgo cluster by
\citet{lee10} reveals that there are numerous GCs wandering in the space
between galaxies.
In NGC 6822, C1 is located well beyond the extent of stellar halo.
Considering that NGC 6822 has no distinct close neighbors,
C1 could be the first candidate of intergalactic GCs in the Local Group.
Future observations to get the velocity of C1 as well as the other ESCs
in NGC 6822 would be invaluable to have better idea on their origins.

\section{Summary and Conclusions}

We have discovered four new star clusters in the halo of dwarf irregular galaxy
NGC 6822 from a wide field imaging survey.
The star clusters have extended structures with $R_{\rm h} \approx 7.5 - 14.0$ pc,
larger than typical GCs by a factor of $2 \sim 5$.
The clusters are distributed over a large space with their projected
galactocentric distance ranging from $10.7 \arcmin$ (about 1.5 kpc) to
$77\arcmin$ (about 11 kpc), far beyond the optical body of NGC 6822.
The spatial distribution of the new star clusters is coincident
with the spatial structure of old stellar halo revealed by recent studies.
The analyses of the integrated color as well as CMDs of cluster member stars suggest
that the new star clusters are likely to be as old as $2 \sim 10$ Gyr, and
the metallicity is estimated to be between Z=0.0001 and Z=0.004 depending
on the adopted age.
Among the four new clusters, at least two clusters, C3 and C4, can be classified
as faint fuzzy clusters based on their very red color with $(V-I)_{0}>1.0$ and
large size with $R_{\rm h}>7.0$ pc. The red color of these two clusters
could be interpreted as the result of moderately enriched metallicity
with Z$\gtrsim 0.004$, which is consistent with the metallicity of existing sample
of faint fuzzy clusters in the literature.

The colors and the sizes of the new clusters are found to be correlated with
the projected distance from the center of NGC 6822 and the clusters located
closer to the center are smaller in size and redder in color than the clusters
in the outer halo.
We suggest that the color gradient may be due to the metallicity gradient
existing in the stellar halo and the differences in the cluster size
may be correlated with the strength of tidal field of NGC 6822: the stronger
the tidal field, the smaller the size of clusters.
This is consistent with the recent theoretical study on the
formation of ESCs, arguing that ESCs
form and survive the disruption only under the weak tidal field condition
that would be possible in small dwarf galaxies like NGC 6822.
However, deeper imaging and high resolution spectroscopic observations are
required to determine the age and the metallicity with higher confidence,
as well as the kinematical properties of these four new ESCs.
The future observational data would be invaluable resources to understand more
clearly
the origin of these ESCs as well as the evolutionary history of
complex dwarf irregular galaxy NGC 6822.

\acknowledgements

The authors are grateful to the anonymous referee for the useful comments
that helped to improve the original manuscript.
N.H. was supported in part by Grant-in-Aid for JSPS Fellow No. 20-08325.
M.G.L. was supported by Mid-career Researcher Program through NRF grant
funded by the MEST (No.2010-0013875).


\end{document}